\documentclass[twocolumn,showpacs,pra,aps,amssymb]{revtex4}
\usepackage{graphicx}
\begin{document}
\title{\bf Quantification of entanglement via uncertainties}

\author{Alexander A. Klyachko, Bar{\i}\c{s} \"{O}ztop, and Alexander S. Shumovsky}

\affiliation{Faculty of Science, Bilkent University, Bilkent,
Ankara, 06800 Turkey}

\begin{abstract}
We show that entanglement of pure multi-party states can be
quantified by means of quantum uncertainties of certain basic
observables through the use of measure that has been initially
proposed in \cite{Baris} for bipartite systems.
\end{abstract}


\maketitle

Recent success in realization of quantum key distribution  has
been achieved through the use of quantum correlations between the
parts in two-qubit systems, which are peculiar to entangled states
(see \cite{Gisin,Zeilinger,Ouelette} and references therein).
Further development of practical implementations of quantum
information technologies requires sources of robust entangled
states and reliable methods of detection of the amount of
entanglement carried by those states (e.g., see
\cite{Hayashi,Enk-2006} and references therein).

There is no universal measure of entanglement suitable for all
systems even in the case of pure states. For example, entanglement
of two qubits is measured by means of {\it concurrence}
\cite{Wootters} for both pure and mixed states. In the case of
pure states, definition of concurrence has been extended to the
bipartite systems with any dimension of the single-party Hilbert
state $d \geq 2$ \cite{Buzek,Mintert}. At the same time, this
definition does not work for systems with number of parts larger
than two. In particular, concurrence is incapable of measuring the
three-party entanglement in three-qubit systems
\cite{Wootters-three-qubits}.

In our previous paper \cite{Baris}, we have find a new
representation of concurrence valid for pure states of an
arbitrary bipartite system which coincides with the Wootters
concurrence~\cite{Wootters} for the case pure two-qubit states. A
logical advantage of this representation is that it expresses
amount of entanglement in terms of variances (quantum
uncertainties) of certain observables. In a sense, this reflects
physical nature of entanglement as manifestation of quantum
uncertainties at their extreme
\cite{Klyachko-2002,Can-2002,Kl-Sh-2004,Kl-Sh-2006} (see also
discussion in Refs. \cite{Guhne,Hofmann,Korepin}).

The main objective of this note is to prove validity of the
measure of Ref. \cite{Baris} for pure states in general settings.

The paper is organized as follows. We start by giving a definition
of basic observables, specifying a given physical system. We
further connect the notion of total variance with measure of
entanglement. Then, we discuss application of this measure to pure
states of two and three qubits. Further, we briefly consider how
this measure works in the case of mixed states. Finally, in
Appendix A, we put the proof of validity of our measure in general
settings.

{\it Quantum dynamical systems}. ---
An idealized von Neumann approach to quantum mechanics, based on
assumption that all Hermitian operators represent measurable
quantities, was first put into question by Wick, Wightman and
Wigner \cite{WWW} in 1952. Later Robert  Hermann \cite{Hermann}
argued soundly   that the basic principles of quantum mechanics
require that measurable observables should form a Lie algebra
$\mathcal{L}$ of (skew)Hermitian operators acting in Hilbert space
$\mathcal{H}$ of the quantum system in question. We refer to
$\mathcal{L}$ as {\it Lie algebra of observables\/} and to the
corresponding Lie group $G=\exp(i\mathcal{L})$ as {\it dynamical
symmetry group\/} of the quantum system.

Restrictions on available observations are of fundamental
importance for physics in general, and for quantum information
specifically. The latter case usually deals with correlated states
of a quantum system with macroscopically separated spatial
components, where only {\it local measurements} are feasible. For
example, the dynamical group of bipartite system
$\mathcal{H}=\mathcal{H}_A\otimes\mathcal{H}_B$ with full access
to local degrees of freedom  amounts to
$\mathrm{SU}(\mathcal{H}_A)\times\mathrm{SU}(\mathcal{H}_B)$.
Without such restrictions, the dynamical group
$G=\mathrm{SU}(\mathcal{H})$ would act transitively on pure states
$\psi\in \mathcal{H}$, which makes them all equivalent. In this
case there would be no place for entanglement and other subtle
quantum phenomena based on intrinsic differences between quantum
states.

{\it Total variance}. --- Recall that uncertainty of an observable
$X \in \mathcal{L}$ in state $\psi \in \mathcal{H}$ is given by
the variance
\begin{eqnarray}
V(X,\psi)=\langle \psi|X^2|\psi \rangle-\langle \psi|X|\psi
\rangle^2. \label{variance}
\end{eqnarray}
Let's now choose an orthonormal basis $X_\alpha$ of the algebra of
observables $\mathcal{L}$ with respect to its {\it Cartan-Killing
form\/} $(X,Y)_K$ \cite{Vinberg2} and define {\it total
variance\/} by equation
\begin{equation}\label{total-variance}
\mathbb{V}(\psi)=\sum_\alpha (\langle \psi|X_\alpha^2|\psi
\rangle-\langle \psi|X_\alpha|\psi \rangle^2).
\end{equation}
For example, for two-qubit system
$\mathcal{H}_A\otimes\mathcal{H}_B$ one can take basis of
$\mathcal{L}=\frak{su}(\mathcal{H}_A)+\frak{su}(\mathcal{H}_B)$,
consisting of Pauli operators $\sigma^A_i$ and $\sigma^B_j$ that
act in components $A$ and $B$, respectively. For a general
multipartite system, the sum (\ref{total-variance}) is extended
over orthonormal bases of traceless local operators for all
parties of the system.

The total variance (\ref{total-variance}) can be understood  as
trace of the quadratic form
$$Q(X)=\langle\psi|X^2|\psi\rangle-\langle\psi|X|\psi\rangle^2,
\quad X\in\mathcal{L}$$
on Lie algebra $\mathcal{L}$, and therefore it is independent of
the basis $X_\alpha$. It measures overall level of quantum
fluctuations of the system in state $\psi$.

The first sum in the total variance (\ref{total-variance})
contains {\it Casimir operator\/} $C=\sum_\alpha X_\alpha^2$,
which acts as a scalar $C_\mathcal{H}$ in every irreducible
representation $G:\mathcal{H}$. As a result we get
\begin{equation}\label{Var_Cas}
\mathbb{V}(\psi)=C_\mathcal{H}-\sum_\alpha\langle
\psi|X_\alpha|\psi \rangle^2.
\end{equation}
To clarify the second sum, consider the average of the basic
observables $X_\alpha$ in state $\psi$
\begin{equation}\label{mean_obs}
X_\psi=\sum_\alpha \langle\psi|X_\alpha|\psi\rangle X_\alpha.
\end{equation}
It can be understood as the {\it center of quantum fluctuations\/}
of the system in state $\psi$. For example, in spin system it is
given by suitably scaled spin projection onto mean spin direction
in state $\psi$. The operator $X_\psi$ is also independent of the
basis $X_\alpha$. This can be seen from the following property
\begin{equation} \label{Kill}\langle\psi|X|\psi\rangle=
(X,X_\psi)_K,\quad \forall
X\in\mathcal{L},
\end{equation}
which holds for basic observables $X=X_\alpha$ by orthogonality
$(X_\alpha,X_\beta)_K=\delta_{\alpha\beta}$, and hence by
linearity for all $X\in\mathcal{L}$. Since Killing form is
nondegenerate, equation (\ref{Kill}) uniquely determines $X_\psi$
and provides for it a coordinate free definition. We show in
Appendix A that the operator $X_\psi$ is closely related to
orthogonal projection of $\rho=|\psi\rangle\langle\psi|$ into Lie
algebra $\mathcal{L}$. The operator $X_\psi$ allows to recast the
total variance (\ref{total-variance}) into the form
\begin{equation}
\mathbb{V}(\psi)=C_\mathcal{H}-\langle\psi|X_\psi|\psi\rangle.
\end{equation}
In Appendix A, we explain how the total variance can be calculated
and give an explicit formula for multi-component system
$\mathcal{H}=\bigotimes_A\mathcal{H}_A$ with full access to local
degrees of freedom in terms of reduced states $\rho_A$
\begin{equation}
 \mathbb{V}(\psi)=\sum_A \left[\dim\mathcal{H}_A-
 \mathrm{Tr}_{\mathcal{H}_A}(\rho_A^2)\right].
 \end{equation}

{\it Completely entangled states}. --- We can infer from
(\ref{Var_Cas}) the inequality
\begin{equation}
\mathbb{V}(\psi)\le C_\mathcal{H}
\end{equation}
which turns into equation iff
\begin{equation}\label{ent_eqn}
\langle\psi|X|\psi\rangle=0,\quad \forall X\in \mathcal{L}.
\end{equation}
For multi-party systems $\mathcal{H}=\bigotimes_A\mathcal{H}_A$,
the latter equation means that all one-party reduced states are
completely disordered. In other words, there exists some local
basis such the the reduced state is given by a diagonal matrix
$\rho_A$, corresponding to uniform probability distribution (that
is, $\rho_A$ are scalar operators). This is a well known
characterization of maximally entangled states. In general we
refer to (\ref{ent_eqn}) as {\it entanglement equation\/} and call
the corresponding state $\psi$ {\it completely entangled.\/}

The completely entangled states are characterized by maximality of
the total variance. Therefore one may be tempted  to consider
entanglement as a manifestation of quantum fluctuations in a state
where they come to their extreme. Entanglement equation
(\ref{ent_eqn}) just states that, in completely entangled state
$\psi$, the quantum system is at the center of its quantum
fluctuations, that is $X_\psi=0$.

{\it Measure of entanglement}. --- States opposite to entangled
ones, to wit those with minimal total level of quantum
fluctuations $\mathbb{V}(\psi)$, for a long time were known as
{\it coherent states\/} \cite{Delbourgo} (see also Refs.
\cite{Klyachko-2002,Viola}). For multi-component systems like
$\mathcal{H}_A\otimes\mathcal{H}_B$ coherent states are just {\it
decomposable\/} or {\it unentangled\/} states
$\psi=\psi_A\otimes\psi_B$.

Observe \cite{Baris} that square of the concurrence $C(\psi)$ for
two component system coincides with the total variance
$\mathbb{V}(\psi)$ reduced to the interval $[0,1]$
\begin{equation}
C^2(\psi)=\frac{\mathbb{V}(\psi)-\mathbb{V}_\mathrm{coh}}
{\mathbb{V}_\mathrm{ent}-\mathbb{V}_\mathrm{coh}},
\end{equation}
where $\mathbb{V}_\mathrm{ent}$ and $\mathbb{V}_\mathrm{coh}$ are
the total level of quantum fluctuations in completely entangled
and coherent states respectively. This clarifies physical meaning
of the concurrence as a measure of overall quantum fluctuations in
the system and leads us to the natural {\it measure of
entanglement\/} of pure states \cite{Baris}
\begin{eqnarray}
\mu(\psi)=
\sqrt{\frac{\mathbb{V}(\psi)-\mathbb{V}_\mathrm{coh}}
{\mathbb{V}_\mathrm{ent}-\mathbb{V}_\mathrm{coh}}}
\label{Concurrence-Baris}
\end{eqnarray}
valid for an arbitrary quantum system. It coincides with the
concurrence for two component systems, but we refrain  to use this
term in general, to avoid confusion with other multicomponent
versions of this notion introduced in \cite{Mintert-2}. We explain
how this measure can be calculated in Appendix A. For a
multicomponent system $\mathcal{H}=\bigotimes_A\mathcal{H}_{A}$,
it can be expressed via local data, encoded in reduced states
$\rho_A$
\begin{equation}\label{mu_rho}
\mu^2(\psi)=\frac{\sum_A\left(1-\mathrm{Tr}\,\rho_A^2\right)}
{\sum_A\left(1-\frac{1}{\dim\mathcal{H}_A}\right)}.
\end{equation}
For example, in two component system $\mathcal{H}=\mathcal{H}_A
\otimes \mathcal{H}_B$ the reduced states $\rho_A$ and $\rho_B$
are isospectral. Hence $\mathrm{Tr}\rho^2_A=\mathrm{Tr}\rho^2_B$
and for system of square format $d\times d$ we arrive at the
familiar formula for concurrence  \cite{Buzek}
\begin{eqnarray}
C(\psi)= \sqrt{\frac{d}{d-1} (1-\mathrm{Tr}\,\rho^2_A)},
\label{Concurrence}
\end{eqnarray}
(in \cite{Buzek} the normalization factor is left adjustable). The
isospectrality of single-party reduced states means that
entanglement can be measured {\it locally}. For example, in the
case of bipartite spin-$s$ system, measurement of only three
observables (spin operators for either party) completely specifies
concurrence (see also discussion in \cite{Huelga}).

An important application for the case of two qubits is provided by
the polarization of photon twins (biphotons) that are created by
the type-II down-conversion \cite{Sergienko}. The spin operators
$S_j$ can be associated with the Stokes operators
\begin{eqnarray}
S_x & \sim & (a_H^+a_V+a^+_Va_H)/\sqrt{2}, \nonumber \\
S_y & \sim & i(a_H^+a_V-a^+_Va_H)/\sqrt{2}, \label{Stokes} \\
S_z & \sim & a^+_Ha_H-a^+_Va_V, \nonumber
\end{eqnarray}
so that the measurement of concurrence (\ref{Concurrence-Baris})
assumes measurement of three Stokes operators for either outgoing
photon beam. Here $a_H$ $(a_V)$ denotes the photon annihilation
operator with horizontal (vertical) polarization. The polarization
of photons is known to be measured by means of either standard
six-state or a minimal four-state ellipsometer \cite{Englert}.

Nevertheless, there is a certain problem with simultaneous
measurement of polarization for one of the two photons created at
once and forming an entangled couple. Because of the commutation
relation
\begin{eqnarray}
[S_j,S_k]=i \epsilon_{jkm}S_m, \quad j,k,m=x,y,z, \nonumber
\end{eqnarray}
the three projections of spin (or three Stokes operators) cannot
be measured independently. The minimal uncertainty relation by
Schr\"{o}dinger \cite{Schrodinger} states
\begin{eqnarray}
V(\psi;S_j)V(\psi;S_k)-(\mathrm{Cov}(S_j,S_k))^2 \nonumber \\ \geq
\frac{1}{4} |\langle \psi|[S_j,S_k]|\psi \rangle|^2 ,
\label{uncertainty-relation}
\end{eqnarray}
where $V(\psi;S_j)$ denotes variance (uncertainty) of observable
$S_j$ in the state $\psi$ and covariance $\mathrm{Cov}(S_j,S_k)$
has the form
\begin{eqnarray}
\mathrm{Cov}(S_j,S_k)= \frac{1}{2} \langle\psi|S_jS_k+S_kS_j|\psi
\rangle - \langle \psi|S_j|\psi \rangle \langle \psi|S_k|\psi
\rangle . \nonumber
\end{eqnarray}
It is straightforward matter to see that the uncertainty relation
is simply reduced to the following one
\begin{eqnarray}
0 \leq \langle\psi|X_{\psi}|\psi\rangle \leq 1/4,
\label{length-limits}
\end{eqnarray}
where $X_{\psi}$ is defined by Eq. (\ref{mean_obs}). Thus, the
uncertainty relation (\ref{uncertainty-relation}) becomes an exact
equality when $\psi = \psi_{coh}$ with
$\langle\psi|X_{\psi}|\psi\rangle=1/4$. In other words, this is an
unentangled biphoton state in which each photon has well-defined
polarization.

In the case of completely entangled biphoton state, the quantity
$\langle\psi|X_{\psi}|\psi\rangle$ has zero value (due to the
condition (\ref{ent_eqn})). In this case, the measurement
performed on a single photon rises an additional question: how to
distinguish between entanglement and classical unpolarized state.

Since Eq. (\ref{length-limits}) is the only relation, connecting
different components of the average spin vector in either party,
the local quantity $\langle\psi|X_{\psi}|\psi\rangle$ cannot be
detected by either single or even two measurements.

{\it Measure $\mu (\psi)$ beyond two-partite states.} ---
Postponing consideration of the measure $\mu(\psi)$ in general
settings till Appendix A, we now note that, in the case of
multipartite system, it gives the total amount of entanglement
carried by all types of inter-party correlations.

For example, the GHZ (Greenberger-Horne-Zeilinger) state of three
qubits
\begin{eqnarray}
|G\rangle = x|000 \rangle +\sqrt{1-|x|^2}|111 \rangle, \quad |x|
\in[0,1], \label{GHZ}
\end{eqnarray}
carries only three-party entanglement. This means that any two
parties are not entangled. In fact, any reduced two-qubit state,
say
\begin{eqnarray}
\rho_{AB}=\mathrm{Tr}_C |G\rangle \langle G|=|x|^2|00 \rangle
\langle 00|+(1-|x|^2)|11 \rangle \langle 11|, \nonumber
\end{eqnarray}
clearly has zero concurrence. The amount of three-part
entanglement in (\ref{GHZ}) is measured by {\it 3-tangle} $\tau$
\cite{Wootters-three-qubits} or {\it Cayley hyperdeterminant}
\cite{Miyake} (for definition of 3-tangle, see Appendix B). It is
easily seen that
\begin{eqnarray}
\tau (G)= \mu^2(G)=4|x|^2(1-|x|^2). \nonumber
\end{eqnarray}
Thus, the squared measure (\ref{Concurrence-Baris}), calculated
for the three-qubit state (\ref{GHZ}), gives the same result as
3-tangle.

Another interesting example is provided by the so-called $W$-state
of three qubits
\begin{eqnarray}
|W \rangle = \frac{1}{\sqrt{3}} (|011\rangle +|010\rangle
+|110\rangle ). \label{W}
\end{eqnarray}
This is a nonseparable state in three-qubit Hilbert space.
Nevertheless, it does not manifest three-party entanglement
because the corresponding 3-tangle $\tau(W)=0$ \cite{Miyake}. At
the same time, the measure (\ref{Concurrence-Baris}) gives
\begin{eqnarray}
\mu(W)= \frac{2\sqrt{2}}{3} \approx 0.94 \label{measure-W}
\end{eqnarray}
because $\mathbb{V}(W)=8+2/3$ and $\mathbb{V}_{\mathrm{coh}}=6$ in
this case. The point is that there is a two-qubit entanglement in
the state (\ref{W}). To justify that the difference $2+2/3$ is
caused just by quantum pairwise correlations, let us calculate the
total {\it covariance}
\begin{eqnarray}
\mathrm{Cov}(W)= \sum_{i=x,y,z} \sum_{J \neq J'} (\langle
W|\sigma_i^J \sigma_i^{J'}|W\rangle \nonumber \\ - \langle
W|\sigma_i^J|W\rangle \langle W|\sigma_i^{J'}|W\rangle ).
\label{Covariance}
\end{eqnarray}
Here $J,J'=A,B,C$ label the parties. It is a straightforward
matter to see that
$\mathbb{V}(W)-\mathbb{V}_{\mathrm{coh}}=\mathrm{Cov}(W)$. Similar
results can be obtained for the so-called {\it biseparable} states
of three qubits
\begin{eqnarray}
(|001\rangle
+|010\rangle),~(|001\rangle+|100\rangle),~(|010\rangle+|100\rangle),
\label{biseparable}
\end{eqnarray}
that also manifest entanglement of two qubits and no entanglement
of all three parts.

Examining entanglement of multi-qubit systems in general (number
of parts is greater than two), it is necessary first to determine
classes of states with different types of entanglement (including
the class of unentangled states). It is assumed that those classes
are nonequivalent with respect to SLOCC (stochastic local
operations assisted by classical communication)~\cite{SLOCC}. The
point is that entanglement of a given type cannot be created or
destroyed under action of SLOCC. In the case of three qubits, such
a classification has been considered in Refs.~\cite{Acin,Miyake}.
In the case of four qubits, the number of classes is much
higher~\cite{Dehaene}. A useful approach to classification is
based on investigation of {\it geometrical invariants} for a given
system (e.g., see Refs.~\cite{Klyachko-2002,Luque}).

For example, the class of four-qubit entangled states can be
specified by the generic GHZ-type state
\begin{eqnarray}
x|0000\rangle \pm \sqrt{1-|x|^2}|1111\rangle,~|x|\in[0,1],
\label{4-qubit-GHZ}
\end{eqnarray}
which becomes completely entangled at $|x|=1/\sqrt{2}$. In
general, four-qubit completely entangled states can be defined by
means of the condition (\ref{ent_eqn}) (see Appendix C). For the
state (\ref{4-qubit-GHZ}), the measure (\ref{Concurrence-Baris})
gives the amount of entanglement $\mu=\sqrt{1-(2|x|^2-1)^2}$,
which becomes complete entanglement at $|x|=1/\sqrt{2}$ as
expected.

At the same time, there is another class of pairwise separable
four-qubit states
\begin{equation}
\frac{1}{\sqrt{2}}(|00\rangle + |11\rangle)\otimes
\frac{1}{\sqrt{2}}(|01\rangle + |10\rangle) ,\nonumber
\end{equation}
in which the first two pairs and the last two pairs separately
manifest complete two-party entanglement, while there is no
four-qubit entanglement (compare with the biseparable states of
three qubits (\ref{biseparable})). In this case, the measure
(\ref{Concurrence-Baris}) again gives the total amount of
entanglement carried by the parts of the system.

{\it Mixed entanglement.\/} --- The measure
(\ref{Concurrence-Baris}) cannot be directly applied to
calculation of entanglement of mixed states because it is
incapable of separation of classical and quantum contributions
into the total variance (\ref{total-variance}). Therefore, $\mu
(\rho)$ always gives estimation from above for the entanglement of
mixed states. This can be easily checked for some characteristic
states like Werner state \cite{Werner} and the so-called maximally
entangled mixed state of Ref. \cite{Kwiat}.

As far as we know, nowadays there is now universally recognized
protocol for separation of classical and quantum uncertainties in
mixed states except the case of two qubits \cite{Wootters}. A
promising approach proposed in Refs. \cite{Mintert,Mintert-2}
consists in the representation of concurrence of a mixed state
$\rho$ as $\mathrm{inf} \sum_i C(\psi_i)$ of all properly
normalized states $\psi$ such that $\rho= \sum_i |\psi_i \rangle
\langle \psi_i|$.

\section*{Summary}

We have shown that description of entanglement in a given system
requires pre-definition of basic observables and that the
entanglement of pure states can be adequately quantified in terms
of total variance of all basic observables. Unlike conventional
concurrence and 3-tangle, that measure the amount of entanglement
of different groups of correlated parties, our measure gives the
total amount of multipartite entanglement, carried by a given
state. Other evident virtues of the measure
(\ref{Concurrence-Baris}) are its simple physical meaning, its
applicability beyond bipartite systems, and its operational
character caused by measurement of quantum uncertainties of
well-defined physical observables.

At the same time, this measure cannot be directly applied to
calculation of entanglement in mixed states. However, it may be
used in the way that has been discussed in Refs.
\cite{Mintert,Mintert-2} as follows
\begin{eqnarray}
\mu (\rho) =\inf \sum_i \mu (\psi_i). \nonumber
\end{eqnarray}

\section*{Acknowledgement}

The authors thank Dr. S.J. van Enk, Dr. V. Korepin and Dr. L.
Viola for useful discussions and indication of their important
works. One of the authors (B. \"{O}.) would like to acknowledge
the Scientific and Technical Research Council of Turkey
(T\"UB\.ITAK) for financial support.

\section*{Appendix A}

Here we calculate the total variance $\mathbb{V}(\psi)$ and the
entanglement measure $\mu(\psi)$.

Let $\mathrm{Herm}(\mathcal{H})$ be space of all Hermitian
operators acting in Hilbert space $\mathcal{H}$ with trace metric
$\mathrm{Tr}_{\mathcal{H}}(XY)$. For {\it simple algebra\/}
$\mathcal{L}$ restriction of the trace metric onto $\mathcal{L}$
is proportional to the Cartan-Killing form
$$\mathrm{Tr}_\mathcal{H}(XY)=D_\mathcal{H}\cdot(X,Y)_K,\quad X,Y\in
\mathcal{L}$$ with the coefficient $D_\mathcal{H}$ known as {\it
Dynkin index\/}
 \cite{Vinberg2}. Consider now orthogonal projection
$\rho_{\mathcal{L}}$
 of $\rho:=|\psi\rangle\langle\psi|
\in \mathrm{Herm}(\mathcal{H})$ into subalgebra $\mathcal{L}
\subset \mathrm{Herm}(\mathcal{H})$, so that
$\mathrm{Tr}_{\mathcal{H}}(\rho
X)=\mathrm{Tr}_{\mathcal{H}}(\rho_{\mathcal{L}}X)$, $\forall
X\in\mathcal{L}$. The projection $\rho_\mathcal{L}$ is closely
related to the mean operator (\ref{mean_obs})
\begin{eqnarray}X_\psi&=&\sum_\alpha
\mathrm{Tr}_\mathcal{H}(\rho X_\alpha) X_\alpha =\sum_\alpha
\mathrm{Tr}_\mathcal{H}(\rho_\mathcal{L} X_\alpha) X_\alpha\nonumber\\
&=&D_\mathcal{H}\sum_\alpha (\rho_\mathcal{L}, X_\alpha)_K
X_\alpha=D_\mathcal{H}\cdot\rho_\mathcal{L}.\nonumber
\end{eqnarray}
Therefore
$$\langle\psi|X_\psi|\psi\rangle=\mathrm{Tr}_\mathcal{H}(\rho
X_\psi)=\mathrm{Tr}_\mathcal{H}(\rho_\mathcal{L}X_\psi)=
D_\mathcal{H}\mathrm{Tr}_\mathcal{H}(\rho_\mathcal{L}^2)$$ and the
total variance (\ref{total-variance}) can be written in the form
\begin{eqnarray}\label{var_smpl}
\mathbb{V}(\psi)=C_\mathcal{H}-\langle\psi|X_\psi|\psi\rangle=
C_\mathcal{H}-D_\mathcal{H}
\cdot\mathrm{Tr}_\mathcal{H}(\rho_\mathcal{L}^2).
\end{eqnarray}
For simple algebra the Casimir $C_{\mathcal{H}}$ and Dynkin index
$D_{\mathcal{H}}$ are given by equations
\begin{equation}\label{Cas_Dyn}
C_{\mathcal{H}}=(\lambda,\lambda+2 \delta), \quad D_{\mathcal{H}}=
\frac{ \dim \mathcal{H}}{\dim \mathcal{L}} (\lambda,\lambda+2
\delta),
\end{equation}
where $\lambda$ denotes the highest weight of irreducible
representation $\mathcal{H}$ and $2 \delta$ is the sum of positive
roots of $\mathcal{L}$. For example, for full algebra of traceless
Hermitian operators $\mathcal{L}=\frak{su}(\mathcal{H})$ we have
\begin{equation}\label{Cas_SU}
C_\mathcal{H}=\dim\mathcal{H}-\frac{1}{\dim\mathcal{H}},\quad
D_\mathcal{H}=1.
\end{equation}

In general, algebra $\mathcal{L}$ splits into simple components
$\mathcal{L}= \bigoplus_A \mathcal{L}_A$ and its irreducible
representation $\mathcal{H}$ into tensor product
$\mathcal{H}=\bigotimes_A \mathcal{H}_A$. In this case equation
(\ref{var_smpl}) should be modified as follows
\begin{eqnarray}\label{var_gnrl}
\mathbb{V}(\psi)= \sum_A \left[C_{\mathcal{H}_A}-D_{\mathcal{H}_A}
\cdot\mathrm{Tr}_{\mathcal{H}_A}(\nu_A^2\rho_{\mathcal{L}_A}^2)\right],
\end{eqnarray}
where $\nu_A=\dim\mathcal{H}/\dim\mathcal{H}_A$.

In quantum information setting $\mathcal{L}_A$ is the full algebra
of traceless Hermitian operators
$X_A:\mathcal{H}_A\rightarrow\mathcal{H}_A$. In this case
everything can be done explicitly.

By definition of reduced states $\rho_A$ we have
\begin{eqnarray}
\mathrm{Tr}_{\mathcal{H}}(\rho X_A)=
\mathrm{Tr}_{\mathcal{H}_A}(\rho_AX_A)
=\nu_A^{-1}\mathrm{Tr}_{\mathcal{H}}(\rho_AX_A). \nonumber
\end{eqnarray}

Comparing this with equation $\mathrm{Tr}_{\mathcal{H}}(\rho
X_A)=\mathrm{Tr}_{\mathcal{H}}(\rho_{\mathcal{L}_A}X_A)$, $\forall
X_A\in \mathcal{L}_A$ characterizing the projection
$\rho_{\mathcal{L}_A}\in\mathcal{L}_A$ we infer
\begin{eqnarray}
\rho_{\mathcal{L}_A}= \nu_A^{-1}\rho^0_A, \nonumber
\end{eqnarray}
where $\rho_A^0=\rho_A-\frac{1}{\dim\mathcal{H}_A}\mathbf{I}$ is
traceless part of $\rho_A$. This allows to calculate the trace

$$\mathrm{Tr}_{\mathcal{H}_A}(\rho_{\mathcal{L}_A}^2)=
\nu^{-2}\left[\mathrm{Tr}_{\mathcal{H}_A}(\rho_A^2)
-\frac{1}{\dim\mathcal{H}_A}\right].$$

Plugging this into equation (\ref{var_gnrl}) and using
(\ref{Cas_SU}) we finally get
 \begin{equation}\label{Var_final}
 \mathbb{V}(\psi)=\sum_A \left[\dim\mathcal{H}_A-
 \mathrm{Tr}_{\mathcal{H}_A}(\rho_A^2)\right].
 \end{equation}
 As an example, consider completely entangled state $\psi$ for which
 $\rho_A=\frac{1}{\dim\mathcal{H}_A}\mathbf{I}$. This gives the
 maximum of the total variance
 $$\mathbb{V}_{\text{max}}=\mathbb{V}_{\text{ent}}=\sum_A \left(
\dim\mathcal{H}_A-\frac{1}{\dim\mathcal{H}_A}\right).$$ The
minimum of the total variance is attained for coherent
(=separable) state $\psi$, for which reduced states $\rho_A$ are
pure. Hence
$$\mathbb{V}_{\text{min}}=\mathbb{V}_{\text{coh}}=\sum_A(\dim\mathcal{H}_A-1).$$
Combining these equations we can write down our measure of
entanglement (\ref{Concurrence-Baris}) explicitly for a
multicomponent system $\mathcal{H}=\bigotimes_A\mathcal{H}_A$ of
arbitrary format
\begin{equation}\label{Amu_rho}
\mu^2(\psi)=\frac{\sum_A[1-\mathrm{Tr}(\rho_A^2)]}
{\sum_A\left(1-\frac{1}{\dim\mathcal{H}_A}\right)}.
\end{equation}

\section*{Appendix B}

For an arbitrary normalized state of three qubits
\begin{eqnarray}
|\psi \rangle =\sum_{\ell,m,n=0}^1 \psi_{\ell m n} |\ell m n
\rangle \nonumber
\end{eqnarray}
the 3-tangle has the form \cite{Wootters-three-qubits,Miyake}
\begin{eqnarray}
\tau (\psi) = 4|\psi_{000}^2 \psi_{111}^2+ \psi_{001}^2
\psi_{110}^2 + \psi_{010}^2 \psi_{101}^2 + \psi_{100}^2
\psi_{011}^2 \nonumber \\
-2(\psi_{000} \psi_{001} \psi_{110} \psi_{111} +\psi_{000}
\psi_{010} \psi_{101} \psi_{111} \nonumber \\  + \psi_{000}
\psi_{100} \psi_{011} \psi_{111} + \psi_{001} \psi_{010}
\psi_{101} \psi_{110} \nonumber \\ + \psi_{001} \psi_{100}
\psi_{011} \psi_{110} + \psi_{010} \psi_{100} \psi_{011}
\psi_{101} ) \nonumber \\ +4( \psi_{000} \psi_{011} \psi_{101}
\psi_{110}  + \psi_{001} \psi_{010} \psi_{100} \psi_{111})|.
\nonumber
\end{eqnarray}

\section*{Appendix C}

A general pure state of four qubits can be written in the form
\begin{equation}
|\psi \rangle = \sum_{k,\ell ,m,n =0}^1 \psi_{k \ell mn} |k,\ell ,
m,n \rangle \label{4-qubit}
\end{equation}
with the normalization condition $\sum_{k,\ell ,m,n =0}^1 |\psi_{k
\ell mn}|^2=1$. Thus, there are 31 real parameters, defining any
state. Condition (\ref{ent_eqn}) gives twelve equations for the
coefficients $\psi_{k \ell mn}$ in (\ref{4-qubit})
\begin{widetext}
\begin{eqnarray*}
  \langle\sigma_{x}^{(A)}\rangle & = & (\psi_{0000}^{*} \psi_{1000} +
  \psi_{0100}^{*} \psi_{1100} + \psi_{0010}^{*} \psi_{1010} +
  \psi_{0001}^{*} \psi_{1001} + \psi_{0110}^{*} \psi_{1110} +
  \psi_{0101}^{*} \psi_{1101}\\
  & & + \psi_{0011}^{*} \psi_{1011} +
  \psi_{0111}^{*} \psi_{1111}) + (c. c.)=0 , \nonumber\\
  \langle\sigma_{x}^{(B)}\rangle &=& (\psi_{0000}^{*} \psi_{0100} +
  \psi_{1000}^{*} \psi_{1100} + \psi_{0010}^{*} \psi_{0110} +
  \psi_{0001}^{*} \psi_{0101} + \psi_{1010}^{*} \psi_{1110} +
  \psi_{1001}^{*} \psi_{1101}\\
  & & + \psi_{0011}^{*} \psi_{0111} +
  \psi_{1011}^{*} \psi_{1111}) + (c. c.)=0 , \nonumber\\
  \langle\sigma_{x}^{(C)}\rangle &=& (\psi_{0000}^{*} \psi_{0010} +
  \psi_{1000}^{*} \psi_{1010} + \psi_{0100}^{*} \psi_{0110} +
  \psi_{0001}^{*} \psi_{0011} + \psi_{1100}^{*} \psi_{1110} +
  \psi_{1001}^{*} \psi_{1011}\\
  & & + \psi_{0101}^{*} \psi_{0111} +
  \psi_{1101}^{*} \psi_{1111}) + (c. c.)=0 , \nonumber\\
  \langle\sigma_{x}^{(D)}\rangle &=& (\psi_{0000}^{*} \psi_{0001} +
  \psi_{1000}^{*} \psi_{1001} + \psi_{0100}^{*} \psi_{0101} +
  \psi_{0010}^{*} \psi_{0011} + \psi_{1100}^{*} \psi_{1101} +
  \psi_{1010}^{*} \psi_{1011}\\
  & & + \psi_{0110}^{*} \psi_{0111} +
  \psi_{1110}^{*} \psi_{1111}) + (c. c.)=0 , \nonumber\\
  \langle\sigma_{y}^{(A)}\rangle & = & i(\psi_{1000}^{*} \psi_{0000} +
  \psi_{1100}^{*} \psi_{0100} + \psi_{1010}^{*} \psi_{0010} +
  \psi_{1001}^{*} \psi_{0001} + \psi_{1110}^{*} \psi_{0110} +
  \psi_{1101}^{*} \psi_{0101}\\
  & & + \psi_{1011}^{*} \psi_{0011} +
  \psi_{1111}^{*} \psi_{0111}) + (c. c.)=0 , \nonumber\\
  \langle\sigma_{y}^{(B)}\rangle &=& i(\psi_{0100}^{*} \psi_{0000} +
  \psi_{1100}^{*} \psi_{1000} + \psi_{0110}^{*} \psi_{0010} +
  \psi_{0101}^{*} \psi_{0001} + \psi_{1110}^{*} \psi_{1010} +
  \psi_{1101}^{*} \psi_{1001}\\
  & & + \psi_{0111}^{*} \psi_{0011} +
  \psi_{1111}^{*} \psi_{1011}) + (c. c.)=0 , \nonumber\\
  \langle\sigma_{y}^{(C)}\rangle &=& i(\psi_{0010}^{*} \psi_{0000} +
  \psi_{1010}^{*} \psi_{1000} + \psi_{0110}^{*} \psi_{0100} +
  \psi_{0011}^{*} \psi_{0001} + \psi_{1110}^{*} \psi_{1100} +
  \psi_{1011}^{*} \psi_{1001}\\
  & & + \psi_{0111}^{*} \psi_{0101} +
  \psi_{1111}^{*} \psi_{1101}) + (c. c.)=0 , \nonumber\\
  \langle\sigma_{y}^{(D)}\rangle &=& i(\psi_{0001}^{*} \psi_{0000} +
  \psi_{1001}^{*} \psi_{1000} + \psi_{0101}^{*} \psi_{0100} +
  \psi_{0011}^{*} \psi_{0010} + \psi_{1101}^{*} \psi_{1100} +
  \psi_{1011}^{*} \psi_{1010}\\
  & & + \psi_{0111}^{*} \psi_{0110} +
  \psi_{1111}^{*} \psi_{1110}) + (c. c.)=0 , \nonumber\\
  \langle\sigma_{z}^{(A)}\rangle & = & |\psi_{0000}|^2 -
  |\psi_{1000}|^2 + |\psi_{0100}|^2 + |\psi_{0010}|^2 +
  |\psi_{0001}|^2 - |\psi_{1100}|^2 - |\psi_{1010}|^2 -
  |\psi_{1001}|^2 + |\psi_{0110}|^2\\
  & & + |\psi_{0101}|^2 +
  |\psi_{0011}|^2 - |\psi_{1011}|^2 - |\psi_{1101}|^2 -
  |\psi_{1110}|^2 + |\psi_{0111}|^2 - |\psi_{1111}|^2=0 , \nonumber\\
  \langle\sigma_{z}^{(B)}\rangle & = & |\psi_{0000}|^2 +
  |\psi_{1000}|^2 - |\psi_{0100}|^2 + |\psi_{0010}|^2 +
  |\psi_{0001}|^2 - |\psi_{1100}|^2 + |\psi_{1010}|^2 +
  |\psi_{1001}|^2 - |\psi_{0110}|^2\\
  & & - |\psi_{0101}|^2 +
  |\psi_{0011}|^2 + |\psi_{1011}|^2 - |\psi_{1101}|^2 -
  |\psi_{1110}|^2 - |\psi_{0111}|^2 - |\psi_{1111}|^2=0 , \nonumber\\
  \langle\sigma_{z}^{(C)}\rangle & = & |\psi_{0000}|^2 +
  |\psi_{1000}|^2 + |\psi_{0100}|^2 - |\psi_{0010}|^2 +
  |\psi_{0001}|^2 + |\psi_{1100}|^2 - |\psi_{1010}|^2 +
  |\psi_{1001}|^2 - |\psi_{0110}|^2\\
  & & + |\psi_{0101}|^2 -
  |\psi_{0011}|^2 - |\psi_{1011}|^2 + |\psi_{1101}|^2 -
  |\psi_{1110}|^2 - |\psi_{0111}|^2 - |\psi_{1111}|^2=0 , \nonumber\\
  \langle\sigma_{z}^{(D)}\rangle & = & |\psi_{0000}|^2 +
  |\psi_{1000}|^2 + |\psi_{0100}|^2 + |\psi_{0010}|^2 -
  |\psi_{0001}|^2 + |\psi_{1100}|^2 + |\psi_{1010}|^2 -
  |\psi_{1001}|^2 + |\psi_{0110}|^2\\
  & & - |\psi_{0101}|^2 -
  |\psi_{0011}|^2 - |\psi_{1011}|^2 - |\psi_{1101}|^2 +
  |\psi_{1110}|^2 - |\psi_{0111}|^2 - |\psi_{1111}|^2=0 , \label{4-qub-ent}
\end{eqnarray*}
\end{widetext}
where
$\langle\sigma_{\alpha}^{(i)}\rangle=\langle\psi_{\mathrm{ent}}
|\sigma_{\alpha}^{(i)}|\psi_{\mathrm{ent}}\rangle$ and $c. c.$
denotes complex conjugate. Thus, there are infinitely many
completely entangled states and the state (\ref{4-qubit-GHZ}) at
$|x|=1/ \sqrt{2}$ is among them.

\end{document}